\documentclass[preprint, aps, prd, nofootinbib, floatfix]{revtex4-1}
\usepackage{graphicx} % Required for inserting images
\usepackage{amsmath, amssymb, amsthm, bm}
\usepackage{hyperref}
\usepackage{array}
\usepackage{mathtools}
\usepackage{xcolor}

\newcommand{\FIRSTAFF}{\affiliation{Department of Physics, University at Buffalo, Buffalo, NY 14260, USA}}

\begin{document}
\title{The BGV Theorem and the Null Convergence Condition}
\author{William H. Kinney}
\email[Electronic address: ]{whkinney@buffalo.edu}
\FIRSTAFF
\date{\today}

\begin{abstract}
We examine the relationship between the Null Convergence Condition (NCC) and the Borde-Guth-Vilenkin (BGV) Theorem. We first show that, for an expanding spacetime foliated orthogonally to a timelike geodesic congruence with vanishing shear and vorticity, the BGV Theorem follows when the NCC holds and the spatial curvature is non-positive, ${}^3\mathcal{R} \leq 0$. The situation becomes more complex in the presence of shear, or non-geodesic threading of the spacetime. In these cases, the local expansion that enters the BGV construction depends not only on the local scalar expansion, but acquires terms given by the contraction of the shear and acceleration with locally defined spatial unit vectors. In the general case, null convergence and non-positive curvature are no longer sufficient to guarantee that the BGV Theorem holds. We discuss the result in the context of eternal inflation in the presence of comoving curvature perturbations and state a more general version of the BGV condition relevant to asymptotically past-de Sitter eternal inflation. 
\end{abstract}

\maketitle

\section{Introduction}

The Borde-Guth-Vilenkin (BGV) Theorem~\cite{Borde:2001nh} is a powerful result linking the local expansion of a cosmological spacetime to past-geodesic completeness. The basic premise of the theorem is that any spacetime with net positive expansion, suitably defined, must be geodesically incomplete. One notable feature of the theorem is that it is purely geometric, and does not rely directly on the assumption of any energy condition. All it requires is positive local expansion along a particular geodesic. However, this says nothing about the physical conditions required to guarantee the correct expansion conditions for the theorem to hold; it is reasonable to expect that this will map to the assumption of certain energy conditions. For example, in a flat Friedmann-Robertson-Walker space, a Hubble parameter satisfying $H > 0$ and ${\dot H} \leq 0$ is related via the Einstein Equations to a perfect fluid which satisfies the Null Energy Condition $\rho + p \geq 0$, which in turn is a result of the Null Convergence Condition $R_{\mu \nu} k^\mu k^\nu \geq 0$, where $R_{\mu \nu}$ is the Ricci tensor, and $k^\mu$ is an arbitrary null geodesic. 

In this paper, we examine more closely the physical conditions necessary for the assumptions of the BGV Theorem to be satisfied, focusing on the Null Convergence Condition (NCC). For a hypersurface-orthogonal, shear-free geodesic perfect-fluid congruence with positive expansion on a final slice, the NCC together with ${}^3\mathcal{R} \leq 0$ provides a uniform positive lower bound on the expansion along past-directed geodesics, and the BGV conclusion follows. We then consider the inclusion of shear, and show that the NCC and non-positive curvature remain sufficient to guarantee that the local scalar expansion $\Theta \equiv u^\mu{}_{;\mu}$ satisfies ${\dot \Theta} \leq 0$ even in the presence of shear. However, the definition of local expansion used in BGV does not depend only on the scalar expansion, but becomes a directional quantity, depending on the shear tensor contracted with locally defined spatial unit vectors. Finally, we consider shear combined with a non-geodesic threading of the spacetime and show that the four-acceleration of the threading must be included in the definition of the local expansion relevant for BGV. In this case, because the term containing the four-acceleration can take either sign, the NCC and non-positive curvature conditions no longer in general guarantee that BGV holds. This latter case is relevant in the presence of comoving curvature perturbations, because comoving world lines are non-geodesic at linear order in perturbation theory.  We discuss this in the context of eternal inflation in the presence of curvature perturbations, and show that in spacetimes that are asymptotically past-de Sitter, a perturbative version of BGV still holds. However, in chaotic eternal inflation, a perturbative construction of BGV is inconclusive. 

\section{The BGV Theorem} \label{sec:BGVShearFree}

Take a timelike geodesic congruence $\left\lbrace u^\mu \right\rbrace$ with zero shear and vorticity $\sigma_{\mu \nu} = \omega_{\mu \nu} = 0$. Then choose a geodesic $\mathcal{C}$ with four-velocity $\left\lbrace v^\mu\right\rbrace \neq \left\lbrace u^\mu\right\rbrace$. We can define
\begin{equation}
    \gamma \equiv - v^\mu u_\mu,
\end{equation}
where $\gamma$ can be identified as the Lorentz factor of the vector $v^\mu$ relative to the rest frame $u^\mu = (1, 0, 0, 0)$. If $v^\mu$ is timelike, its normalization is
\begin{equation} \label{eq:normalization}
    \begin{aligned}
        v^\mu v_\mu &= -1\\
                    &= v^\mu \left(g_{\mu \sigma} v^\sigma\right) \\
                    &= v^\mu \left(\lambda_{\mu \sigma} - u_\mu u_\sigma\right) v^\sigma \\
                    &= - \gamma^2 + \lambda_{\mu \nu} v^\mu v^\nu. 
    \end{aligned}
\end{equation}
Here $\lambda_{\mu \nu}$ is the spatial projection tensor,
\begin{equation}
    \lambda_{\mu \nu} \equiv g_{\mu \nu} + u_\mu u_\nu,\quad \lambda_{\mu \nu} u^\nu = 0. 
\end{equation}
Normalization of $v^\mu$ therefore results in the relation
\begin{equation}
    \lambda_{\mu \nu} v^\mu v^\nu = \gamma^2 - 1. 
\end{equation}
We also have the geodesic equation,
\begin{equation}
    \frac{d v^\mu}{d s} = v^\nu v^\mu{}_{;\nu} = 0,
\end{equation}
where $ds$ is the proper time measured along $v^\mu$. Then
\begin{equation} \label{eq:dgammads}
    \begin{aligned}
        \frac{d \gamma}{d s} &= - \frac{d}{d s} \left(v^\mu u_\mu\right)\\
                             &= - v^\mu \frac{d u_\mu}{d s} \\
                             &= - v^\mu v^\nu u_{\mu;\nu} = - \frac{1}{3} \Theta v^\mu v^\nu \lambda_{\mu \nu} \\
                             &= \frac{1}{3} \Theta \left(1 - \gamma^2\right),
    \end{aligned}
\end{equation}
where the expansion rate $\Theta$ relative to $\left\lbrace u^\mu \right\rbrace$ is defined by the divergence of the four-velocity
\begin{equation}
    \Theta \equiv u^\mu{}_{;\mu}. 
\end{equation} 
We then have the relation
\begin{equation}
    \Theta = \frac{3}{1 - \gamma^2} \frac{d \gamma}{d s}.
\end{equation}
Integrating with respect to proper time along the geodesic $\mathcal{C}$,
\begin{equation} \label{eq:FTimelike}
    \begin{aligned}
        \int{\Theta ds} &= \int{\frac{3}{1 - \gamma^2} \frac{d \gamma}{d s} ds} \\
                        &= \int{\frac{3 d\gamma}{1 - \gamma^2}} \\
                        &= \frac{3}{2} \ln\left(\frac{\gamma + 1}{\gamma - 1}\right).
    \end{aligned}
\end{equation}
Similarly, for a null geodesic $k^\mu$, define
\begin{equation}
    \gamma \equiv - k^\mu u_\mu.
\end{equation}
Here $\gamma$ does not have the interpretation of a Lorentz factor as in the timelike case. Normalization implies
\begin{equation}
    k^\mu k_\mu = 0 = - \gamma^2 + \lambda_{\mu \nu} k^\mu k^\nu,
\end{equation}
and the spatial projection is
\begin{equation}
    \lambda_{\mu \nu} k^\mu k^\nu = \gamma^2. 
\end{equation}
We can write the geodesic equation for $k^\mu$ in terms of an affine parameter $\alpha$ as
\begin{equation}
    \frac{d k^\mu}{d \alpha} = k^\nu k^\mu{}_{;\nu} = 0. 
\end{equation}
For the null case, the derivative of $\gamma$ with respect to the affine parameter satisfies
\begin{equation} \label{eq:gammanull}
    \begin{aligned}
        \frac{d \gamma}{d \alpha} &= - \frac{d}{d \alpha} \left(k^\mu u_\mu\right)\\
                             &= - k^\mu \frac{d u_\mu}{d \alpha} \\        
                             &= - k^\mu k^\nu u_{\mu;\nu} = - \frac{1}{3} \Theta k^\mu k^\nu \lambda_{\mu \nu} \\
                             &= - \frac{1}{3} \gamma^2 \Theta.
    \end{aligned}
\end{equation}
As in the timelike case, we can then write the expansion parameter as
\begin{equation}
    \Theta = - \frac{3}{\gamma^2} \frac{d \gamma}{d \alpha},
\end{equation}
and
\begin{equation} \label{eq:FNull}
    \begin{aligned}
        \int{\Theta d\alpha} &= - \int{\frac{3}{\gamma^2} \frac{d \gamma}{d \alpha} d\alpha} \\
                        &= -\int{\frac{3 d\gamma}{\gamma^2}} \\
                        &= \frac{3}{\gamma}. 
    \end{aligned}
\end{equation}

The original formulation of the BGV Theorem in Ref.~\cite{Borde:2001nh} contains loopholes, which we discuss below. A rigorous statement of the theorem that avoids those loopholes is as follows~\cite{Pavlovic:2023mke,Garcia-Saenz:2024ogr}:
\newtheorem*{BGV}{BGV Theorem}
\begin{BGV}
    Consider a geodesic $\mathcal{C}$ parameterized by affine parameter $\alpha$. Let $\left(\alpha_i,\alpha_f\right)$ be the maximal past-directed affine domain of the geodesic $\mathcal{C}$. If there exists a constant $\Delta > 0$ such that
    \begin{equation} 
        \Theta_{\text{av}} \equiv \frac{1}{\alpha_f - \alpha_0} \int_{\alpha_0}^{\alpha_f}{\Theta d\alpha} \geq \Delta\quad \forall \alpha_0 \in \left(\alpha_i,\alpha_f\right),
    \end{equation}
    then $\mathcal{C}$ is past-incomplete on the interval $\left(\alpha_i,\alpha_f\right)$. 
\end{BGV}

\begin{proof}
    We use $\alpha$ as a general affine parameter, which reduces to proper time $s$ for timelike geodesics and is affine for null geodesics. From Eqs. (\ref{eq:FTimelike}) and (\ref{eq:FNull}), 
    \begin{equation} 
        \Theta_{\text{av}} \equiv \frac{1}{\alpha_f - \alpha_0} \int_{\alpha_0}^{\alpha_f}{\Theta d\alpha} = \frac{F\left(\gamma_f\right) - F\left(\gamma_0\right)}{\alpha_f - \alpha_0} \geq \Delta,
    \end{equation}
    where $F\left(\gamma\right)$ is given by 
    \begin{equation}
        F\left(\gamma\right) = \frac{3}{2} \ln\left(\frac{\gamma + 1}{\gamma - 1}\right),
    \end{equation}
    if $\mathcal{C}$ is timelike, and 
    \begin{equation}
        F\left(\gamma\right) = \frac{3}{\gamma},
    \end{equation}
    if $\mathcal{C}$ is null. Since $F\left(\gamma\right) > 0$ for both timelike and null geodesics,
    \begin{equation}
        \Delta \leq \Theta_{\text{av}} < \frac{F\left(\gamma_f\right)}{\alpha_f - \alpha_0}\quad \forall \alpha_0 \in \left(\alpha_i,\alpha_f\right).
    \end{equation}
    Then
    \begin{equation}
        \alpha_0 > \alpha_f - \frac{F\left(\gamma_f\right)}{\Delta}\quad \forall \alpha_0 \in \left(\alpha_i,\alpha_f\right).
    \end{equation}
    Therefore $\alpha_i$ is bounded from below,
    \begin{equation}
        \alpha_i \geq \alpha_f - \frac{F\left(\gamma_f\right)}{\Delta},
    \end{equation}
    and the affine length of the geodesic $\mathcal{C}$ is finite. Therefore the spacetime is geodesically incomplete on the interval. 
\end{proof}

Two qualifications concerning the interpretation of the BGV result are worth noting. First, it is possible to construct a monotonically expanding universe which evades the BGV Theorem, and is geodesically complete. A simple example is the ``loitering'' universe, 
\begin{equation} \label{eq:Loitering}
    a\left(t\right) = a_i \left(1 + e^{t / \tau}\right),\quad \tau = \text{const.} >0.
\end{equation}
Unlike the de Sitter case, the scale factor asymptotically approaches a finite value $a \to a_i$ as $t \to -\infty$, and the geometry approaches Minkowski space at early times. The spacetime is past geodesically complete despite being monotonically expanding~\cite{Ellis:2002we,Lesnefsky:2022fen}. Such loitering models can evade the BGV bound precisely because their early-time behavior asymptotically mimics Minkowski space. To see why it is not a counterexample, consider the integral
\begin{equation} 
    \int_{-\infty}^{t_0}{H dt} = \ln{\left(1 + e^{t_0 / \tau}\right)}. 
\end{equation}
This is finite and positive. However, since the interval in question is infinite as $\Delta t \equiv t_0 - t_i \to \infty$, the average Hubble parameter vanishes for any $t_0$:
\begin{equation} 
   H_{\text{av}} = \lim_{t_i \to -\infty} \frac{1}{\Delta t} \int_{t_i}^{t_0}{H d t} = 0. 
\end{equation}
Contrast this with the case of de Sitter evolution
\begin{equation}
    a\left(t\right) \propto e^{H t},\quad H = \text{const.}.
\end{equation}
In the de Sitter case,
\begin{equation} \label{eq:Hav}
    H_{\text{av}} = \lim_{t_i \to -\infty} \frac{1}{\Delta t} \int_{t_i}^{t_0}{H dt} =  H > 0.
\end{equation}
The average (\ref{eq:Hav}) along the interval is finite and positive, and the spacetime is geodesically incomplete. In a flat background, the scale factor (\ref{eq:Loitering}) also has ${\dot H} > 0$, and therefore violates the Null Energy Condition, since ${\dot H} > 0$ requires $p < -\rho$. This suggests that the Null Energy Condition $p \geq -\rho$ may provide a dynamical condition for maintaining the positive expansion required by the BGV bound. We consider the relation between the BGV bound and the Null Energy Condition in the next section. 

In a second class of extensions, the geodesically incomplete coordinate patch is completed by embedding it in a larger spacetime across its past boundary. Constructions by Aguirre and Gratton~\cite{Aguirre:2001ks,Aguirre:2003ck} and Ferreira \textit{et al.}~\cite{Ferreira:2025rje} are of this type. A simple example of this is the flat patch on de Sitter space; the past boundary of the flat de Sitter chart is a regular null boundary of that chart, and the chart embeds as a proper subset of the geodesically complete global de Sitter manifold. BGV correctly diagnoses the incompleteness of the flat patch; it does not establish that the boundary is a curvature singularity or that the patch is inextendible. The flat patch can therefore be extended across its past boundary to the full de Sitter hyperboloid, which is geodesically complete. This highlights an important property of the BGV Theorem: BGV can only tell you that a spacetime is incomplete. It does not tell you the spacetime is singular, or inextendible. The general criteria for determining extendibility remain an open question~\cite{Geshnizjani:2023hyd}.

\section{The Null Convergence Condition and BGV}

\subsection{Example: FRW Spacetime}

We begin with the simple example of an FRW spacetime. Take $\Theta = 3 H$, and the Friedmann Equation is
\begin{equation} \label{eq:FriedmannEquation}
    \frac{1}{3} \Theta^2 = 3 H^2 = 8 \pi G \rho - \frac{3 k}{a^2}. 
\end{equation}
Then the Raychaudhuri Equation is
\begin{equation}
    3 {\dot H} + 3 H^2 = -R_{\mu \nu} u^\mu u^\nu = -4 \pi G \left(\rho + 3 p\right). 
\end{equation}
Choosing the normalization $u_\mu k^\mu = -1$, we obtain
\begin{equation}
    \begin{aligned}
        {\dot \Theta} = 3 {\dot H} &= -4 \pi G \left(\rho + 3 p\right) - 3 H^2 \\
        &= -12 \pi G \left(\rho + p\right) + \frac{3 k}{a^2} \\
        &= -\frac{3}{2} R_{\mu \nu} k^\mu k^\nu + \frac{3 k}{a^2}.
    \end{aligned}
\end{equation}
We can write this in terms of the intrinsic curvature of spatial hypersurfaces,
\begin{equation}
    {}^3\mathcal{R} = \frac{6 k}{a^2},
\end{equation}
so that 
\begin{equation}
    {\dot \Theta} = -\frac{3}{2} R_{\mu \nu} k^\mu k^\nu + \frac{1}{2} \left({}^3\mathcal{R}\right). 
\end{equation}
Therefore, the Null Convergence Condition 
\begin{equation}
    R_{\mu \nu} k^\mu k^\nu \geq 0,
\end{equation}
combined with
\begin{equation}
    {}^3\mathcal{R} \leq 0,
\end{equation}
implies
\begin{equation}
    {\dot \Theta} \leq 0.
\end{equation}

An example of the positive-curvature case is closed coordinates on de Sitter space. The Friedmann Equation is
\begin{equation}
    H^2 = \left(\frac{\dot a}{a}\right)^2 = \frac{\Lambda}{3} - \frac{1}{a^2},
\end{equation}
with solution
\begin{equation}
    a\left(t\right) = \sqrt{\frac{3}{\Lambda}} \cosh\left(\sqrt{\frac{\Lambda}{3}} t\right),
\end{equation}
so that the Hubble parameter is
\begin{equation}
    H = \sqrt{\frac{\Lambda}{3}} \tanh\left({\sqrt{\frac{\Lambda}{3}} t}\right).
\end{equation}
The universe undergoes a nonsingular bounce, and is geodesically complete.

\subsection{General Perfect Fluids}

We next generalize to a perfect fluid without assuming an FRW background. Take a timelike congruence $u^\mu$ which is geodesic and hypersurface-orthogonal,
\begin{equation}
    {\dot u^\mu} = u^\nu u^\mu{}_{;\nu} = 0. 
\end{equation}
Any null vector $k^\mu$ can be expressed in terms of the timelike congruence $u^\mu$ and a vector $e^\mu$ as
\begin{equation}
    k^\mu = C \left(u^\mu + e^\mu\right),
\end{equation}
where
\begin{equation}
    u^\mu e_\mu = 0. 
\end{equation}
Then
\begin{equation}
    k^\mu k_\mu = 0 = C^2 \left(u^\mu u_\mu + e^\mu e_\mu\right) \Rightarrow e^\mu e_\mu = +1,
\end{equation}
that is, the vector $e^\mu$ is a unit spacelike four-vector. We can take $C = 1$ without loss of generality. Then
\begin{equation}
    R_{\mu \nu} k^\mu k^\nu = R_{\mu \nu} u^\mu u^\nu + 2 R_{\mu \nu} u^\mu e^\nu + R_{\mu \nu} e^\mu e^\nu. 
\end{equation}
We take the stress energy to be
\begin{equation}
    \begin{aligned}
        T_{\mu \nu} &= \rho u_\mu u_\nu + p \lambda_{\mu \nu} \\
        &= \left(\rho + p\right) u_\mu u_\nu + p g_{\mu \nu}. 
    \end{aligned}
\end{equation}
Then the trace of the stress-energy is
\begin{equation}
    T^{\lambda}{}_{\lambda} = \left(\rho + p\right) u^\lambda u_\lambda + p g^\lambda{}_\lambda = -\rho + 3 p. 
\end{equation}
For a perfect fluid, the Ricci tensor $R_{\mu \nu}$ can be related to the stress-energy via the Einstein Field Equation
\begin{equation}
    \begin{aligned}
        R_{\mu \nu} &=  8 \pi G \left(T_{\mu \nu} - \frac{1}{2} g_{\mu \nu} T^\mu{}_{\mu}\right) \\
        &= 8 \pi G\left[\left(\rho + p\right) u_\mu u_\nu + p g_{\mu \nu} + \frac{1}{2} g_{\mu \nu} \left(\rho - 3 p\right)\right] \\
        &= 8 \pi G \left[\left(\rho + p\right) u_\mu u_\nu + \frac{1}{2} \left(\rho - p\right) g_{\mu \nu}\right]. 
    \end{aligned}
\end{equation}
Contracting with a null congruence $k^\mu$ yields 
\begin{equation}
    R_{\mu \nu} k^\mu k^\nu = 8 \pi G \left(\rho + p\right). 
\end{equation}
The Raychaudhuri Equation for a timelike congruence $u^\mu$ is then:
 \begin{equation} \label{eq:RaychaudhuriEquation}
    \dot\Theta + \frac{1}{3} \Theta^2 + \left(\sigma^2 - \omega^2\right) = - R_{\mu \nu} u^\mu u^\nu = -4 \pi G \;\left(\rho + 3 p\right),
\end{equation}
where
\begin{equation}
    \sigma^2 \equiv \sigma^{\mu \nu} \sigma_{\mu \nu},\qquad \omega^2 \equiv \omega^{\mu \nu} \omega_{\mu \nu}.
\end{equation}
We will assume vanishing vorticity, $\omega_{\mu \nu} = 0$, so that the congruence is hypersurface-orthogonal. The Hamiltonian constraint on hypersurfaces orthogonal to the congruence $u^\mu$ is
\begin{equation} \label{eq:ADMHamiltonianConstraint}
    {}^3 \mathcal{R}  - \left(K^{\mu \nu} K_{\mu \nu} - K^2\right) = 16 \pi G \rho,
\end{equation}
where $\rho = T_{\mu \nu} u^\mu u^\nu$ is the matter Hamiltonian, ${}^3\mathcal{R}$ is again the intrinsic curvature of spatial hypersurfaces, and $K_{\mu \nu}$ is the extrinsic curvature, defined as the spatially projected gradient of the four-velocity,
\begin{equation} \label{eq:ExtrinsicCurvature}
    \begin{aligned}
        K_{\mu \nu} &\equiv \lambda_\nu{}^\sigma u_{\mu;\sigma} \\
                    &= u_\mu{}_{;\nu} + u_\nu u^\sigma u_\mu{}_{;\sigma} \\
                    &= u_\mu{}_{;\nu} + {\dot u_\mu} u_\nu \\
                    &= \sigma_{\mu \nu} + \omega_{\mu\nu} + \frac{1}{3} \Theta \lambda_{\mu \nu}.
    \end{aligned}
\end{equation}
For zero vorticity, this tensor is symmetric, and
\begin{equation}
    K \equiv \lambda^{\mu \nu} K_{\mu \nu} = \frac{1}{3} \Theta \lambda^\mu{}_\mu = \Theta,
\end{equation}
while
\begin{equation}
K_{\mu\nu}K^{\mu\nu} = \frac{1}{3}\Theta^2 + \sigma_{\mu\nu}\sigma^{\mu\nu}.
\end{equation}
The Hamiltonian constraint is then
\begin{equation} \label{eq:PerfectFluidHamiltonianConstraint}
    {}^3 \mathcal{R} + \frac{2}{3} \Theta^2 -  \sigma^2 = 16 \pi G \rho.
\end{equation}
This is just the generalization of the Friedmann Equation (\ref{eq:FriedmannEquation}). Substituting into the Raychaudhuri Equation (\ref{eq:RaychaudhuriEquation}) yields
\begin{equation} \label{eq:NECThetadot}
    \begin{aligned}
        {\dot \Theta} &= - 12 \pi G \left(\rho + p\right) + \frac{1}{2} {}^3 \mathcal{R} - \frac{3}{2} \sigma^2 \\
                      &= - \frac{3}{2} R_{\mu \nu} k^\mu k^\nu + \frac{1}{2} {}^3 \mathcal{R} - \frac{3}{2} \sigma^2. 
    \end{aligned}
\end{equation}
Then the Null Convergence Condition
\begin{equation}
    R_{\mu \nu} k^\mu k^\nu \geq 0
\end{equation}
combined with non-positive intrinsic curvature
\begin{equation}
    {}^3 \mathcal{R} \leq 0,
\end{equation}
yields
\begin{equation}
    {\dot \Theta} \leq 0,
\end{equation}
even for the case of nonzero shear, since $\sigma^2$ is non-negative. However, in the presence of shear, the scalar expansion $\Theta$ is no longer the fundamental quantity that appears in the BGV Theorem. We demonstrate this in the next section. 

\section{The BGV Theorem Including Shear}

In Section~\ref{sec:BGVShearFree}, we derived the BGV Theorem assuming shear-free flow. Here we re-derive the theorem including shear, while continuing to assume a geodesic, vorticity-free threading. Defining $v^\mu$ as before, Eq. (\ref{eq:dgammads}) in the presence of shear $\sigma_{\mu \nu}$ becomes
\begin{equation}
    \begin{aligned}
        \frac{d \gamma}{d s} &= - \frac{d}{d s} \left(v^\mu u_\mu\right)\\
                             &= - v^\mu \frac{d u_\mu}{d s} \\
                             &= - v^\mu v^\nu u_{\mu;\nu} \\
                             & = - \frac{1}{3} \Theta v^\mu v^\nu \lambda_{\mu \nu}  - \sigma_{\mu \nu} v^\mu v^\nu \\
                             &= \frac{1}{3} \Theta \left(1 - \gamma^2\right) - \sigma_{\mu \nu} v^\mu v^\nu \\
                             &= \frac{1}{3} \left(1 - \gamma^2\right) \Theta_v,
    \end{aligned}
\end{equation}
where we define the expansion along the geodesic $v^\mu$ as
\begin{equation} \label{eq:DefThetav}
    \Theta_v \equiv  \Theta + \frac{3 \sigma_{\mu \nu} v^\mu v^\nu}{\gamma^2 - 1}. 
\end{equation}
Then Eq. (\ref{eq:FTimelike}) carries through relative to the expansion $\Theta_v$:
\begin{equation} 
    \begin{aligned}
        \int{\Theta_v ds} &= \int{\frac{3}{1 - \gamma^2} \frac{d \gamma}{d s} ds} \\
                        &= \int{\frac{3 d\gamma}{1 - \gamma^2}} \\
                        &= \frac{3}{2} \ln\left(\frac{\gamma + 1}{\gamma - 1}\right).
    \end{aligned}
\end{equation}

The null case is similar: Eq. (\ref{eq:gammanull}) becomes
\begin{equation}
    \begin{aligned}
        \frac{d \gamma}{d \alpha} &= - \frac{d}{d \alpha} \left(k^\mu u_\mu\right)\\
                             &= - k^\mu \frac{d u_\mu}{d \alpha} \\        
                             &= - k^\mu k^\nu u_{\mu;\nu} \\
                             & = - \frac{1}{3} \Theta k^\mu k^\nu \lambda_{\mu \nu}  - \sigma_{\mu \nu} k^\mu k^\nu \\
                             &= - \frac{1}{3} \gamma^2 \Theta  - \sigma_{\mu \nu} k^\mu k^\nu \\
                             &= - \frac{1}{3} \gamma^2 \Theta_k, 
    \end{aligned}
\end{equation}
where
\begin{equation} \label{eq:DefThetak}
    \Theta_k \equiv \Theta + \frac{3 \sigma_{\mu \nu} k^\mu k^\nu}{\gamma^2}.
\end{equation}
As in the timelike case, we can then write the expansion parameter as
\begin{equation}
    \Theta_k = - \frac{3}{\gamma^2} \frac{d \gamma}{d \alpha},
\end{equation}
and
\begin{equation} 
    \begin{aligned}
        \int{\Theta_k d\alpha} &= - \int{\frac{3}{\gamma^2} \frac{d \gamma}{d \alpha} d\alpha} \\
                        &= -\int{\frac{3 d\gamma}{\gamma^2}} \\
                        &= \frac{3}{\gamma}. 
    \end{aligned}
\end{equation}

We can interpret $\Theta_v$ and $\Theta_k$ geometrically by defining spatial unit vectors by
\begin{equation} \label{eq:Defemuv}
    e^\mu_v \equiv \frac{\lambda^\mu{}_\nu v^\nu}{\sqrt{\gamma^2 - 1}},\qquad e^\mu_v e_{v \mu} = +1
\end{equation}
for the timelike geodesic tangent $v^\mu$, so that
\begin{equation}
    v^\mu = \gamma u^\mu + \sqrt{\gamma^2 - 1} e^\mu_v. 
\end{equation}
Similarly, for the null vector $k^\mu$, we define
\begin{equation} \label{eq:Defemuk}
    e^\mu_k \equiv \frac{\lambda^\mu{}_\nu k^\nu}{\gamma},\qquad e^\mu_k e_{k \mu} = +1,
\end{equation}
and
\begin{equation}
    k^\mu = \gamma \left(u^\mu + e^\mu_k\right). 
\end{equation}
Then Eq. (\ref{eq:DefThetav}) becomes
\begin{equation}
    \Theta_v = \Theta + 3 \sigma_{\mu \nu} e^\mu_v e^\nu_v,
\end{equation}
and Eq. (\ref{eq:DefThetak}) is
\begin{equation}
    \Theta_k = \Theta + 3 \sigma_{\mu \nu} e^\mu_k e^\nu_k.
\end{equation}
We then see that the generalized expansion is \textit{directional}, due to the presence of shear. The proof of the BGV Theorem proceeds exactly as in Sec.~\ref{sec:BGVShearFree}, with the scalar expansion $\Theta$ replaced by the directional expansion $\Theta_v$ or $\Theta_k$. 

From Eq. (\ref{eq:NECThetadot}), the Null Convergence Condition combined with ${}^3\mathcal R\leq0$ is sufficient to ensure $\dot\Theta\leq0$. However, this does not by itself guarantee positive directional expansion, since
\begin{equation}
    \Theta_e = \Theta + 3\sigma_{\mu\nu} e^\mu e^\nu
\end{equation}
depends explicitly on the direction of the shear. The Null Convergence Condition and non-positive intrinsic curvature control the evolution of the scalar expansion $\Theta$, but do not themselves guarantee the same behavior for $\Theta_e$. In the presence of shear, a separate geometric constraint on the directional expansion is therefore required.

\section{Non-geodesic Threading}

We next consider the case for which the congruence $u^\mu$ defining the threading of the spacetime is non-geodesic,
\begin{equation}
    \dot u^\mu = u^\nu u^\mu{}_{;\nu} \neq 0. 
\end{equation}
This is relevant, for example, in linear perturbation theory. Comoving world lines in FRW space are geodesic at background order, but are non-geodesic at linear order. Conversely, choosing a geodesic threading introduces an effective gauge-dependent anisotropic stress $T_{i j} \neq 0,\quad i \neq j$ at linear order; one can have either a geodesic threading, or vanishing anisotropic stress, but not both (we discuss this more fully in Sec.~\ref{sec:Discussion}). 

Defining the expansion variable relevant to the BGV construction proceeds in very much the same way as the addition of shear. For a non-geodesic congruence, the velocity gradient is
\begin{equation}
    u_{\mu;\nu}
    =
    \sigma_{\mu\nu}
    + \omega_{\mu\nu}
    + \frac{1}{3}\Theta\lambda_{\mu\nu}
    - {\dot u}_\mu u_\nu.
\end{equation} 
Then for a timelike geodesic with tangent $v^\mu$,
\begin{equation}
    \begin{aligned}
        \frac{d \gamma}{d s} &= - v^\mu v^\nu u_{\mu;\nu} \\
                             &= \frac{1}{3} \Theta \left(1 - \gamma^2\right) - \sigma_{\mu \nu} v^\mu v^\nu - \gamma {\dot u}_\mu v^\mu\\
                             &= \frac{1}{3} \left(1 - \gamma^2\right) {\tilde \Theta}_v,
    \end{aligned}
\end{equation}
where we define the generalized expansion parameter
\begin{equation}
    \begin{aligned}
        {\tilde \Theta}_v &\equiv \Theta + \frac{3 \sigma_{\mu \nu} v^\mu v^\nu}{\gamma^2 - 1} + \frac{3 \gamma {\dot u}_\mu v^\mu}{\gamma^2 - 1} \\
                          &= \Theta + 3 \sigma_{\mu \nu} e^\mu_v e^\nu_v + \frac{3 \gamma}{\sqrt{\gamma^2 - 1}} {\dot u}_\mu e^\mu_v \\
                          &= \Theta + 3 \sigma_{\mu \nu} e^\mu_v e^\nu_v + \frac{3}{\beta} {\dot u}_\mu e^\mu_v.
    \end{aligned}
\end{equation}
The spacelike unit vector $e^\mu_v$ is again defined as in Eq. (\ref{eq:Defemuv}), and
\begin{equation}
    \beta \equiv \frac{\sqrt{\gamma^2 - 1}}{\gamma}
\end{equation}
is the usual relative speed. The BGV integral retains its original form:
\begin{equation}
    \int{{\tilde \Theta}_v ds} = \frac{3}{2} \ln\left(\frac{\gamma + 1}{\gamma - 1}\right).
\end{equation}
Similarly, for null geodesics,
\begin{equation}
    \begin{aligned}
        \frac{d \gamma}{d \alpha} &= - k^\mu k^\nu u_{\mu;\nu} \\
                                  &= - \frac{1}{3} \gamma^2 \Theta  - \sigma_{\mu \nu} k^\mu k^\nu  - \gamma {\dot u}_\mu k^\mu\\
                                &= - \frac{1}{3} \gamma^2 {\tilde \Theta_k}, 
    \end{aligned}
\end{equation}
where
\begin{equation}
    \begin{aligned}
        {\tilde \Theta}_k &\equiv \Theta + \frac{3 \sigma_{\mu \nu} k^\mu k^\nu}{\gamma^2} + \frac{3 {\dot u}_\mu k^\mu}{\gamma} \\
                 &= \Theta + 3 \sigma_{\mu \nu} e^\mu_k e^\nu_k + 3 {\dot u}_\mu e^\mu_k,
    \end{aligned}
\end{equation}
where the spacelike unit vector $e^\mu_k$ is again defined as in Eq. (\ref{eq:Defemuk}). The BGV integral is
\begin{equation}
    \int{{\tilde \Theta_k} d \alpha} = \frac{3}{\gamma}. 
\end{equation}
The corresponding observer-dependent expansion rate was derived by Kothawala~\cite{Kothawala:2018bgv}. Here we have decomposed it explicitly into scalar expansion, shear, and acceleration.

We can then write a fully general BGV condition for a timelike geodesic tangent $v^\mu$ as
\begin{equation}
    \frac{1}{s_f - s_0} \int_{s_0}^{s_f}{\left(\Theta + 3 \sigma_{\mu \nu} e^\mu_v e^\nu_v + \frac{3}{\beta} {\dot u}_\mu e^\mu_v \right) ds} \geq \Delta > 0\qquad \forall s_0 \in \left(s_i, s_f\right). 
\end{equation}
This need not be satisfied globally; geodesic incompleteness requires only that the BGV condition hold along a single past-directed geodesic. 

A non-geodesic threading also significantly weakens the relationship between the Null Convergence Condition and the sign of $\dot \Theta$, since the acceleration divergence ${\dot u}^\mu{}_{;\mu}$ has indefinite sign, and---continuing to assume a perfect fluid with vanishing vorticity---the Raychaudhuri equation is
\begin{equation} 
    {\dot \Theta} = - \frac{3}{2} R_{\mu \nu} k^\mu k^\nu + \frac{1}{2} {}^3 \mathcal{R} - \frac{3}{2} \sigma^2 + {\dot u}^\mu{}_{;\mu}. 
\end{equation}
Once the threading is non-geodesic, the Null Convergence Condition combined with ${}^3 \mathcal{R} \leq 0$ no longer guarantees ${\dot \Theta} \leq 0$. 

\section{Discussion} \label{sec:Discussion}

We have seen that in the isotropic limit, with geodesic threading $u^\mu$, the BGV integral depends on the local scalar expansion, defined as the divergence of the threading four-velocity $\Theta = u^\mu{}_{;\mu} > 0$. In this limit, the Null Convergence Condition
\begin{equation}
    R_{\mu \nu} k^\mu k^\nu \geq 0,
\end{equation}
combined with non-positive three-curvature of spatial hypersurfaces
\begin{equation}
    {}^3 \mathcal{R} \leq 0
\end{equation}
is sufficient to guarantee $\dot \Theta \leq 0$, and the spacetime is consequently geodesically past-incomplete. However, the inclusion of shear and/or a non-geodesic threading complicates this intuitive picture, since the BGV integral no longer depends only on the local scalar expansion, but includes additional terms depending on the directional shear $\sigma_{\mu \nu}$ and acceleration ${\dot u}^\mu$, 
\begin{equation}
    {\tilde \Theta}_v = \Theta + 3 \sigma_{\mu \nu} e^\mu_v e^\nu_v + \frac{3}{\beta} {\dot u}_\mu e^\mu_v,
\end{equation}
for a timelike geodesic tangent vector
\begin{equation}
    v^\mu = \gamma u^\mu + \sqrt{\gamma^2 - 1} e^\mu_v,
\end{equation}
where $e^\mu_v$ is a unit-normalized spacelike vector orthogonal to $u^\mu$, 
\begin{equation}
    e^\mu_v e_{v \mu} = +1,\qquad e^\mu_v u_\mu = 0,
\end{equation}
so that
\begin{equation}
    v^\mu u_\mu = -\gamma. 
\end{equation}
The corresponding expressions for a null tangent vector $k^\mu$ are
\begin{equation}
    {\tilde \Theta}_k = \Theta + 3 \sigma_{\mu \nu} e^\mu_k e^\nu_k + 3 {\dot u}_\mu e^\mu_k,
\end{equation}
where 
\begin{equation}
    k^\mu = \gamma \left(u^\mu + e^\mu_k\right), 
\end{equation}
with
\begin{equation}
     e^\mu_k e_{k \mu} = +1,\qquad e^\mu_k u_\mu = 0.
\end{equation}
In the presence of acceleration, the Null Convergence Condition and non-positive curvature are no longer sufficient to guarantee ${\dot \Theta} \leq 0$, since
\begin{equation} 
    {\dot \Theta} = - \frac{3}{2} R_{\mu \nu} k^\mu k^\nu + \frac{1}{2} {}^3 \mathcal{R} - \frac{3}{2} \sigma^2 + {\dot u}^\mu{}_{;\mu},
\end{equation}
and ${\dot u}^\mu{}_{;\mu}$ can carry either sign. 

It is useful to contrast this result with the special case of a geodesic comoving perfect fluid with vanishing momentum density. The momentum constraint is
\begin{equation}
    \nabla_\nu \sigma^{\mu\nu}
    - \frac{2}{3}\nabla^\mu\Theta=0,
\end{equation}
where $\nabla_\mu$ is the spatially projected covariant derivative. In the shear-free limit, this implies $\nabla^\mu\Theta=0$, so the scalar expansion is constant on each spatial hypersurface. In this restricted case, the directional dependence of the BGV expansion disappears. More generally, spatial variations in $\Theta$ are related to the divergence of the shear, while the shear and four-acceleration also enter the BGV expansion directly through their projections along the chosen geodesic. The directionality of the generalized expansion is therefore a physical consequence of departing from shear-free geodesic flow.

We are most interested in applying the BGV Theorem to the case of eternal inflation. In the presence of perturbations, comoving world lines are non-geodesic at linear order,
\begin{equation} \label{eq:ComovingGeodesicEq}
    \begin{aligned}
        {\dot u}_0 &= 0,\\
        {\dot u}_i &= \partial_i \left[A + \left(a H\right)B + B'\right] \\
                   &= \partial_i \left[\zeta - \left(a H\right) B\right],
    \end{aligned}
\end{equation}
where $\zeta$ is the curvature perturbation, $A$ is the lapse perturbation $g_{0 0} \equiv - a^2 \left(1 + 2 A\right)$ and $B$ is the shift perturbation, $g_{0 i} \equiv a^2 \partial_i B$. The last line follows from the constraint of vanishing anisotropic stress, $G_{i j} = 8 \pi G T_{i j} = 0,\ i \neq j$. Eternal inflation occurs in the limit where the comoving curvature power spectrum becomes non-perturbative,
\begin{equation}
    P_\zeta^{1/2} = \frac{H^2}{2 \pi \left\vert \dot \phi\right\vert} > 1,
\end{equation}
so the acceleration $\dot u^\mu$ is not necessarily small. In the case of eternal inflation on a plateau potential, such as Starobinsky inflation, the divergence of $\zeta$ as $\dot\phi \rightarrow 0$ is purely a gauge artifact. On the plateau, $V'\left(\phi\right) \rightarrow 0$, so the classical background trajectory also has $\dot\phi \rightarrow 0$, with the background expansion rate $H \rightarrow \text{const.}$ In this limit the comoving curvature perturbation diverges:
\begin{equation}
    \zeta \sim \frac{H^2}{\dot \phi} \rightarrow \infty.
\end{equation}
However, $\zeta$ is not gauge-invariant. The gauge-invariant metric potential is the Mukhanov variable, defined in terms of the comoving curvature perturbation $\zeta$ and the field perturbation $\delta \phi$ as
\begin{equation}
    u = a \left(\delta \phi + \frac{\dot\phi}{H} \zeta\right).
\end{equation}
The Mukhanov variable is well-behaved in the limit $\dot\phi \rightarrow 0$, 
\begin{equation}
    u \rightarrow a \delta\phi,\quad \dot\phi \rightarrow 0. 
\end{equation}
In this limit, comoving gauge becomes singular, while the expansion along past-directed geodesics is asymptotically de Sitter. A regular gauge describing the expansion is then flat gauge, $\zeta \equiv 0$~\cite{Kinney:2005vj}. In flat gauge, the field perturbation $\delta\phi$ decouples from the metric at linear order, and behaves as a free field in a de Sitter background. The field perturbation $\delta \phi$ is dominated by quantum fluctuations, and becomes a stochastic variable. The rms fluctuation at the Hubble scale is
\begin{equation} \label{eq:phiexp}
   \sqrt{\left\langle \delta \phi^2 \right\rangle_{k = a H}} = \frac{H}{2 \pi} = \frac{\Theta}{6 \pi}. 
\end{equation}
Perturbations to the stress-energy are then quadratic in the field perturbations:
\begin{equation}
    \delta T_{\mu \nu} \sim \partial_\mu \left(\delta \phi\right) \partial_\nu \left(\delta \phi\right). 
\end{equation}
Using Eq. (\ref{eq:phiexp}), we then expect perturbations in the local expansion rate to be of order
\begin{equation}
    \frac{\delta {\tilde \Theta}}{\tilde \Theta} \sim \mathcal{O}\left[\left(\frac{\Theta}{M_{\text{P}}}\right)^2\right],
\end{equation}
and similarly perturbations to $\dot \Theta$ are of order
\begin{equation}
    \frac{\delta {\dot \Theta}}{\Theta^2} \sim \mathcal{O}\left[\left(\frac{\Theta}{M_{\text{P}}}\right)^2\right].
\end{equation}
Therefore, perturbations to the generalized local expansion remain small, and the spacetime is asymptotically de Sitter. Corrections to the isotropic BGV integral likewise remain small, and the BGV Theorem can be consistently applied. 

The case of eternal inflation on a chaotic-type potential, for example
\begin{equation}
    V\left(\phi\right) = m^2 \phi^2,
\end{equation}
is considerably less clear. For the chaotic potential, the classical slow-roll solution is
\begin{equation}
    \dot\phi = \text{const.}
\end{equation}
Eternal inflation occurs in the limit that the expansion rate becomes large,
\begin{equation}
    H_{\text{EI}} \gg H_*,
\end{equation}
where $H_*$ is the expansion rate during slow-roll inflation~\cite{Barenboim:2016mmw}. The comoving curvature perturbation again becomes large, $P_\zeta^{1/2} > 1$, leading to eternal inflation. In this case, the field is free to fluctuate arbitrarily high on the potential, as long as the local energy density remains sub-Planckian; field fluctuations can in principle source large fluctuations in the energy density, and therefore the local expansion parameter $\Theta$. In this regime, the stochastic spacetime can no longer be treated as a small perturbation of a single slow-roll background, and the background solution alone no longer controls the realization-dependent directional shear and acceleration terms. Establishing a uniform positive lower bound on ${\tilde \Theta}$ requires a non-perturbative, gauge-invariant treatment of the stochastic geometry. A perturbative application of BGV is therefore inconclusive.

\section*{Acknowledgments}

This work is supported by the National Science Foundation under grant NSF-PHY-2310363. I thank Diego Rios for helpful conversations. 

\bibliography{bibliography}

\end{document}